# Continuous Affect Prediction Using Eye Gaze and Speech

Jonny O' Dwyer, Ronan Flynn, Niall Murray
Department of Electronics and Informatics
Athlone Institute of Technology
Athlone, Ireland
j.odwyer@research.ait.ie, rflynn@ait.ie, nmurray@research.ait.ie

*Abstract* — **Affective computing research traditionally focused on labeling a person's emotion as one of a discrete number of classes e.g. happy or sad. In recent times, more attention has been given to continuous affect prediction across dimensions in the emotional space, e.g. arousal and valence. Continuous affect prediction is the task of predicting a numerical value for different emotion dimensions. The application of continuous affect prediction is powerful in domains involving real-time audio-visual communications which could include remote or assistive technologies for psychological assessment of subjects. Modalities used for continuous affect prediction may include speech, facial expressions and physiological responses. As opposed to single modality analysis, the research community have combined multiple modalities to improve the accuracy of continuous affect prediction. In this context, this paper investigates a continuous affect prediction system using the novel combination of speech and eye gaze. A new eye gaze feature set is proposed. This novel approach uses open source software for real-time affect prediction in audio-visual communication environments. A unique advantage of the human-computer interface used here is that it does not require the subject to wear specialized and expensive eye-tracking headsets or intrusive devices. The results indicate that the combination of speech and eye gaze improves arousal prediction by 3.5% and valence prediction by 19.5% compared to using speech alone.**

## I. INTRODUCTION

Affective computing is the interdisciplinary study of human emotional analysis, synthesis, recognition, and prediction. This field of research combines approaches from a range of disciplines, including computer science, neuroscience, cognitive science and psychology. For example, machine learning techniques are often used to classify a person as happy or sad as guided by the psychological literature on facial expressions associated with these emotions. Within affective computing, emotion recognition is the process of classifying a person's emotional state into one of a number of predefined classes based on observed data [1–4]. Happy, sad, disgust, positive, and negative are examples of emotional classes. Continuous affect prediction is the task of predicting a numerical value for different emotion dimensions, such as arousal or valence [5-6]. The arousal emotion dimension describes the level of energy in an observed emotion and valence describes how positively or negatively

This research was supported by the Irish Research Council grant number GOIP/2016/1572.

pleasurable that emotion is. Figure 1 shows some emotions projected onto the valence-arousal space as presented in [7].

The availability of high quality audio-visual affective computing databases, such as MAHNOB-HCI [4], AVEC 2014 [8], RECOLA [9] and SEMAINE [10], have allowed the affective computing community to further its research. In addition to audio-visual modalities the MAHNOB-HCI [4] and RECOLA [9] databases provide physiological data. These databases are provided with ground-truth annotations for machine predictor or classifier model training, with which unseen test data can be evaluated.

Speech as a modality has been widely used within affective computing as is evident from studies such as [2-6], [11], and [12]. In addition, the availability of speech-based feature sets, for example AVEC 2014 [8], GeMAPS [13] and ComParE [14], and open source tools like openSMILE [15] make speech accessible as an input modality for use by affective computing researchers.

The study of eye gaze for affective computing tasks has received increased attention in recent times with studies such as [1], [4] and [16] attempting emotion classification and [17] performing eye gaze classification (looking or not-looking at stimuli). The open source software tool OpenFace [18] makes eye gaze tracking from video data possible without the need for expensive eye tracking equipment. The AVEC 2016 challenge [19] used eye gaze approximation data obtained using OpenFace as part of the corpus provided for the depression recognition challenge.

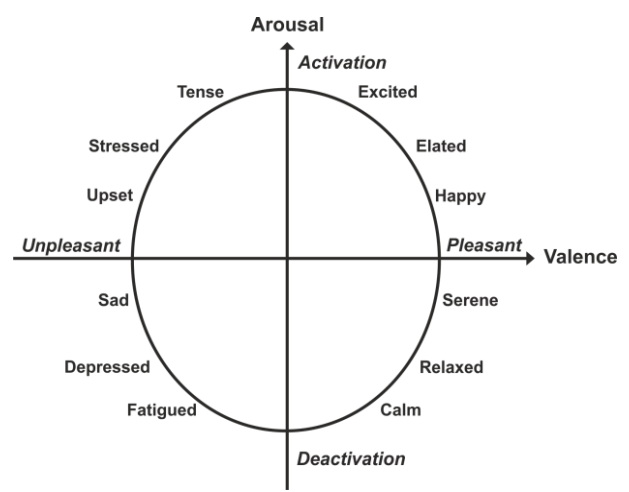

Fig. 1 Arousal-Valence diagram from Abhang and Gawali [7]

To the authors' knowledge, no eye gaze feature set has been applied to the task of continuous affect prediction. The novelty of this work lies in that fact that it demonstrates the benefit of considering eye gaze for continuous affect prediction, particularly in a bimodal system when combined with speech.

When working with multiple input modalities, some form of fusion method must be used to exploit the continuous affect prediction of the individual modalities. Feature-level fusion, which is sometimes referred to as early feature fusion, is where the features extracted from the different modalities are combined into a single, larger feature file that is then used to create a single model for continuous affect prediction. Decision-level fusion, also known as model fusion or late feature fusion, is where prediction models are created for each individual modality and the outputs, or decisions, from these models are then combined to give an overall prediction for the particular affect dimension. Both feature-level and decision-level fusion techniques were investigated in [5], in which the modalities of speech, facial expression and physiological data were used for continuous affect prediction tasks on the RECOLA [9] database. Several neural network architectures were investigated in [5] and from the experiments it was found that long short-term memory recurrent neural networks (LSTM-RNN) performed better than feed-forward neural networks for affect prediction. In addition, decision-level fusion was found to perform best when compared to feature-level fusion. The authors of [6] investigated support vector regression (SVR) and bi-directional long short-term memory recurrent neural network (BLSTM-RNN) machine learning schemes for continuous affect prediction on the SEMAINE corpus [10]. The multimodal input consisted of audio, facial expressions and shoulder gestures for arousal and valence prediction experiments. Based on evaluation results, the higher performing BLSTM-RNN machine learning scheme was selected over the SVR method for input fusion experiments. The fusion techniques investigated in [6] included feature, model and output-associative fusion. The model-level fusion described in [6] was the same as the decision-level fusion method investigated in [5]. The authors of [6] proposed an output-associative fusion framework that was designed to exploit the correlations and co-variances between the arousal and valence emotion dimensions. The output-associative fusion was found in [6] to be the fusion method that performed best in terms of root mean squared error and linear correlation.

This paper proposes a new eye gaze feature set for continuous affect prediction. Raw eye gaze data were extracted using OpenFace [18] from audio-visual data provided with the AVEC 2014 corpus [8]; this was followed by feature extraction from the raw data. The proposed eye gaze feature set is evaluated for continuous arousal and valence prediction using the AVEC 2014 [8] database. A number of different fusion methods to combine the new eye gaze feature set with the AVEC 2014 speech feature set were investigated. The experimental results for arousal and valence prediction clearly demonstrate the benefit of using eye gaze and speech as bimodal inputs to a continuous affect prediction system.

The layout of this paper is as follows: Section 2 presents the eye gaze feature set used in this work. The experimental framework used is detailed in Section 3. Experimental results and discussion of the results are presented in Section 4. Concluding remarks are given in Section 5.

## II. EYE GAZE FEATURE SET

The proposed eye gaze feature set presented here was inspired by [4]. Raw eye gaze data were gathered from video sequences using OpenFace [18], which was followed by feature extraction from the raw data. The full list of features calculated from the raw eye gaze data is detailed in Table I. The proposed set, which consists of 31 features, is the first set using eye gaze cues compiled for continuous affect prediction.

There are a number of key differences between the eye gaze feature set presented in this paper and that presented in [4]. The eye gaze feature set in [4] was used for an emotion classification task in which the arousal and valence of a person was classed into one of three states after the person had watched a video with emotional content. Arousal was classed as low, medium or activated while valence was classed as unpleasant, neutral or pleasant. The eye gaze feature set presented here is intended for continuous affect prediction, that is, the prediction of a numerical value on a continuous basis for both arousal and valence throughout the entire duration of a video, be it live or recorded. A Tobii X120 [20] eye-tracking device, worn by the person under evaluation, was used to obtain the required data for the eye gaze features in [4]. The focus in this work is audio and video data where no specialized hardware is required. OpenFace [18], an open source software tool, was used to obtain the necessary eye data required for the eye gaze feature set proposed here. In terms of the eye gaze feature sets, pupillometry and blink data were used in [4] but not in the feature set proposed here. Such data can easily be captured by the headgear used in [4], which is designed for eye tracking. Eye-close frame count data, which is not used in [4], was found to be of benefit and is included in the proposed eye gaze feature set. Another difference between the feature set presented in this work and [4] is the use of eye scan paths, based on [21], and inter quartile ranges of eye gaze coordinates. The inter quartile ranges of eye gaze coordinates were added to the proposed eye gaze feature set based on experimentation and performance optimization for the intended application of continuous affect prediction from audio and video.

TABLE I.    PROPOSED EYE GAZE FEATURE SET

| Data | Features |
|---|---|
| Eye gaze distance (2 features) | eye gaze approach ratio, average eye gaze approach time in milliseconds |
| Eye scan paths (2 features) | average scan path length, standard deviation of scan path lengths |
| Vertical and horizontal eye gaze coordinates (24 features) | average, inter quartile range 1-2, inter quartile range 2-3, standard deviation, skewness, power spectral densities at frequencies [0.011, 0.022, 0.033-0.044, 0.055-0.066, 0.077-0.133] Hz, average of standard deviation of coordinates in each fixation zone, standard deviation of standard deviation of coordinates in each fixation zone |
| Eye closure (3 features) | average eye close frame count, standard deviation of eye close frame count, skewness of eye close frame count |

III. EXPERIMENTAL FRAMEWORK

This section presents the experimental framework used for a bimodal eye gaze and speech continuous affect prediction system. The database used, feature extraction, machine learning method, and the fusion methods evaluated are all discussed.

*A. Database*

The Freeform task of the AVEC 2014 audio-visual corpus [8] was used for experimental evaluations. The Freeform task provides a training set, development set and test set, equally split over the 150 audio-visual recordings. Annotator-rated ground-truth values for both arousal and valence are provided for the recordings.

*B. Feature Extraction*

Raw eye gaze data was gathered using OpenFace [18] and this was followed by feature extraction. The features were extracted from the eye gaze data using overlapping 3 second video segments, with a 1 second overlap between adjacent segments. This method is the same as the short segmentation method used in [8] and allowed for the seamless alignment of features extracted from both speech and eye gaze. As shown in Table I there were 31 eye gaze features calculated. The features from the speech were extracted using openSMILE [15]. A total of 2,268 speech features were extracted using the AVEC 2014 short segmentation method as described in [8].

*C. Creating Models for Affect Prediction*

The SMOreg function in the WEKA data mining toolkit [22] was used to build models for arousal and valence emotion dimension prediction. The SMOreg function is the WEKA implementation of SVR and was also used to provide baseline results published for AVEC 2014 [8]. Of the training set data, 66% was used for model building with the remaining 34% used for model validation.

The AVEC 2014 dataset contained significantly more 0.0 ground-truth rated valence values from the annotators than 0.0 rated arousal values. It was suggested in [17] that valence might be perceived unconsciously by human raters. To consider the effects of this on the models generated to continuously predict affect, two experiments were carried out. Firstly, the 0.0 ground-truth rated valence values were included in valence model training and secondly, the 0.0 ground-truth values were omitted from valence model training. The removal of 0.0 annotated valence ratings resulted in better valence prediction accuracy for all models except for the feature fusion-based model. Therefore, the exclusion of 0.0 ground-truth valence values was applied to the building of all experimental valence models except for the feature fusion case. The complexity, or *C* values, used to control the SVR bias-variance trade-off for each of the models used in this work can be seen in Table II. The *C* values resulting in the best correlation of predicted values with ground-truth values following a series of experiments were selected for each model.

*D. Fusion Methods*

The feature fusion methods employed here followed those used in [6], with the addition of a simple *averaged prediction fusion* method. Four fusion methods were evaluated in total: feature fusion (early feature fusion), averaged prediction fusion, model fusion (or late feature fusion), and output-associative fusion.

Feature fusion, a technique commonly employed in multimodal affective computing [5], involves the row-wise concatenation of the features from each modality into one larger feature set for each segment. For example, each segment of annotator-rated ground-truth data in this work would have 2,268 AVEC 2014 speech features and 31 eye gaze features for a total feature vector dimensionality of 2,299 combined speech and eye gaze features. This fusion method is illustrated in Figure 2. The full training and development sets provided with AVEC were used as the training set for this work, with 34% of this new and larger training set held back for validation as described under training Set ID A in Table III.

For the averaged prediction fusion method, the predictions of speech and eye gaze for a given segment are averaged to give the final prediction for the segment as shown in Figure 3.

TABLE II. EXPERIMENTAL COMPLEXITY C VALUES

| Model | C |
|---|---|
| Unimodal Speech Arousal | $2.5 \times 10^{-4}$ |
| Unimodal Speech Valence | $9.0 \times 10^{-5}$ |
| Unimodal Eye Gaze Arousal | 0.009 |
| Unimodal Eye Gaze Valence | 6.5 |
| Feature Fusion Arousal | $1.8 \times 10^{-4}$ |
| Feature Fusion Valence | $2.0 \times 10^{-4}$ |
| Speech Model Fusion Arousal | $7.0 \times 10^{-5}$ |
| Speech Model Fusion Valence | $8.0 \times 10^{-6}$ |
| Eye Gaze Model Fusion Arousal | 10.0 |
| Eye Gaze Model Fusion Valence | 10.0 |
| Final Model Fusion Arousal | 9.0 |
| Final Model Fusion Valence | 7.0 |
| Final Output Associative Fusion Arousal | 0.2 |
| Final Output Associative Fusion Valence | $4.0 \times 10^{-4}$ |

TABLE III.  TRAINING SETS USED FOR MODEL BUILDING. IN ALL CASES 34% OF THE TRAINING SET IS USED FOR MODEL VALIDATION

| Model(s) | Arousal and Valence Training Set | Set ID |
|---|---|---|
| All Unimodal | Combined AVEC 2014 Training and Development Sets | A |
| Feature Fusion | Combined AVEC 2014 Training and Development Sets | A |
| Averaged Prediction Fusion | Combined AVEC 2014 Training and Development Sets | A |
| Speech Model Fusion Arousal | AVEC 2014 Training Set | B |
| Speech Model Fusion Valence | AVEC 2014 Training Set | B |
| Eye Gaze Model Fusion Arousal | AVEC 2014 Training Set | B |
| Eye Gaze Model Fusion Valence | AVEC 2014 Training Set | B |
| Final Model Fusion Arousal | Model Fusion Development Set Arousal Predictions | C |
| Final Model Fusion Valence | Model Fusion Development Set Valence Predictions | C |
| Final Output Associative Fusion Arousal | Model Fusion Development Set Arousal and Valence Predictions | E |
| Final Output Associative Fusion Valence | Model Fusion Development Set Arousal and Valence Predictions | E |

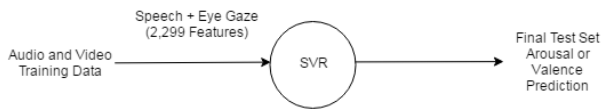

Fig. 2: Speech and eye gaze feature level fusion

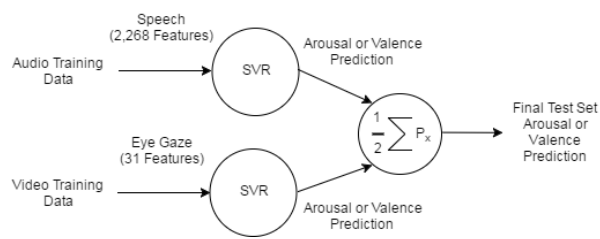

Fig. 3: Averaged prediction fusion

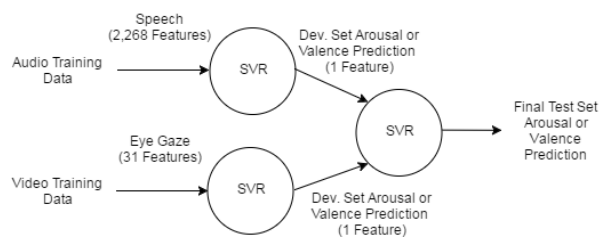

Fig. 4: Model based input fusion

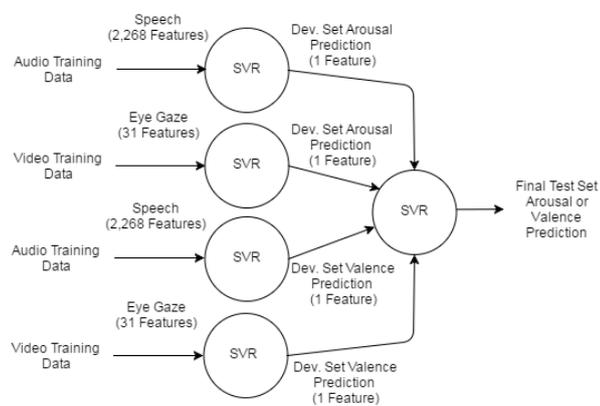

Fig. 5: Output associative based input fusion

For model fusion, SVR prediction models for speech and eye gaze were trained separately on the AVEC 2014 training set. Following this, separate speech and eye gaze predictions were made on the development set. Finally, these speech and eye gaze predictions were combined to form the training feature vector for an additional SVR that produced the final test set prediction model. A diagram illustrating the model fusion process can be seen in Figure 4. The training and validation set configurations for this fusion process are identified as Set ID B and Set ID C in Table III.

Output-associative fusion, which was proposed in [6], aims to exploit the correlations and co-variances between arousal and valence emotion dimensions. As can be seen in Figure 5, both arousal and valence SVR predictions from each modality provide training input to a final SVR to make final predictions. The training and validation scheme for model building and verification was the same as that for the model fusion previously discussed, except for the final prediction model which used arousal and valence predictions together from each modality, identified as Set ID E in Table III.

## IV. RESULTS AND DISCUSSION

The results obtained for the experimental evaluation of the fusion of the eye gaze features proposed in Section 2 with the AVEC 2014 speech features [8] are detailed in Table IV. This table shows the performance achieved for the machine prediction systems on the AVEC 2014 [8] test set. The results are presented in terms of Pearson's correlation coefficient ($r$) and concordance correlation coefficient (CCC) between machine predicted values for arousal and valence versus ground-truth values. These performance metrics allow us to evaluate linear correlation ($r$) and agreement (CCC) between our machine predicted values and that of ground-truth. CCC, which has been used in recent work such as [5] and [19], combines $r$ with the square difference between the mean of the predicted and ground-truth values as in (1). The CCC of machine prediction performance penalises correlated time series that are shifted in value.

$$CCC = \frac{2\rho\sigma_x\sigma_y}{\sigma_x^2 + \sigma_y^2 + (\mu_x - \mu_y)^2} \quad (1)$$

TABLE IV. PREDICTION PERFORMANCE OF UNIMODAL BASELINE SYSTEMS AND BIMODAL FUSED SPEECH AND EYE GAZE SYSTEMS

| Method | Arousal | | Valence | |
|---|---|---|---|---|
| | r | CCC | r | CCC |
| Unimodal Speech | 0.523 | 0.339 | 0.311 | 0.23 |
| Unimodal Eye Gaze | 0.322 | 0.154 | 0.331 | 0.212 |
| Feature Fusion | 0.537 | 0.351 | 0.331 | 0.244 |
| Averaged Prediction Fusion | 0.549 | 0.257 | 0.421 | 0.239 |
| Model Fusion | 0.509 | 0.264 | 0.335 | 0.275 |
| Output Associative Fusion | 0.526 | 0.296 | 0.347 | 0.241 |

For the unimodal systems considered, Table IV shows that the best arousal prediction correlation was 0.5225 for the AVEC 2014 speech feature set. This observation shows that the speech signal is correlated with emotional arousal to a greater degree than eye gaze from video. In addition, the results in Table IV indicate that a speech input to the SVR system resulted in more accurate arousal prediction with a CCC value of 0.339 for unimodal speech compared to 0.154 for unimodal eye gaze. Given the smaller feature vector size of eye gaze (N = 31) compared with that of speech modality (N = 2,268), the eye gaze performance for valence prediction is significant. From Table IV it can be seen that unimodal eye gaze SVR predictions were better correlated with valence ($r$ = 0.331) compared with speech ($r$ = 0.311). In addition, eye gaze performed only slightly worse in terms of CCC achieving a score of 0.212 compared with a value of 0.23 for speech.

The results from Table IV show that only one of the four proposed bimodal speech and eye gaze systems outperformed the unimodal speech system for arousal prediction agreement as measured by CCC. For valence, all four bimodal systems showed increased levels of agreement in terms of CCC when compared to either the speech or eye gaze unimodal systems. The best prediction performance for arousal and valence in terms of CCC were 0.351 (feature fusion) and 0.275 (model fusion) respectively. These scores represent a 3.5% improvement for arousal and a 19.5% improvement for valence compared to the best unimodal performances of the speech-based system.

From the unimodal experiments, eye gaze does not appear to perform well for arousal prediction. The trajectory of the speech and eye gaze feature fusion system's arousal predictions can be seen in Figure 6 where it can be observed that feature fusion follows the trajectory of the actual, or ground-truth rated, values for arousal only marginally closer than speech alone. All bimodal systems presented provided an increase in prediction accuracy for valence over unimodal speech. This, in addition to the unimodal eye gaze system's comparable performance to speech for valence prediction is evidence that eye gaze when considered from video is indicative of level of valence. The highest performing fusion method for valence, model-based speech and eye gaze fusion, can be seen plotted in Figure 7. Figure 7 shows a much larger difference between unimodal speech predictions and that of the fusion method, with the model fusion method's trajectory more closely matching that of ground-truth valence values. From these experiments, the benefit of combining eye gaze features from video with speech is clear, in particular for valence dimension prediction.

## V. CONCLUSIONS

This paper has presented a new eye gaze feature set for use in a continuous affect prediction system. Experimental results show the benefit of combining eye gaze with speech for use in a bimodal continuous affect prediction system. Considering eye gaze and speech together in a bimodal system yields a 19.5% improvement in valence prediction and a 3.5% improvement in arousal prediction compared to speech alone. The additional eye gaze features used in the experiments are only 1.37% of the speech feature set size. The results demonstrate the benefit of using eye gaze in a multi-modal continuous affect prediction system and confirm that further investigation into the use of eye gaze for affect prediction should be carried out.

A potential area of application for the proposed system is in remote health diagnostics where data pertaining to the emotions of the subject may be useful. A significant advantage of the human-computer interface in the proposed system is its simplicity: it uses open source software and there is no intrusive hardware to be worn by persons interacting with the system.

Future work will include further enhancement of the proposed eye gaze feature set, attribute selection to reduce the total number of speech and eye gaze features, and continuous affect prediction using the proposed bimodal system on other audio-visual corpora. Additional machine learning schemes such as LSTM-RNN will also be investigated.

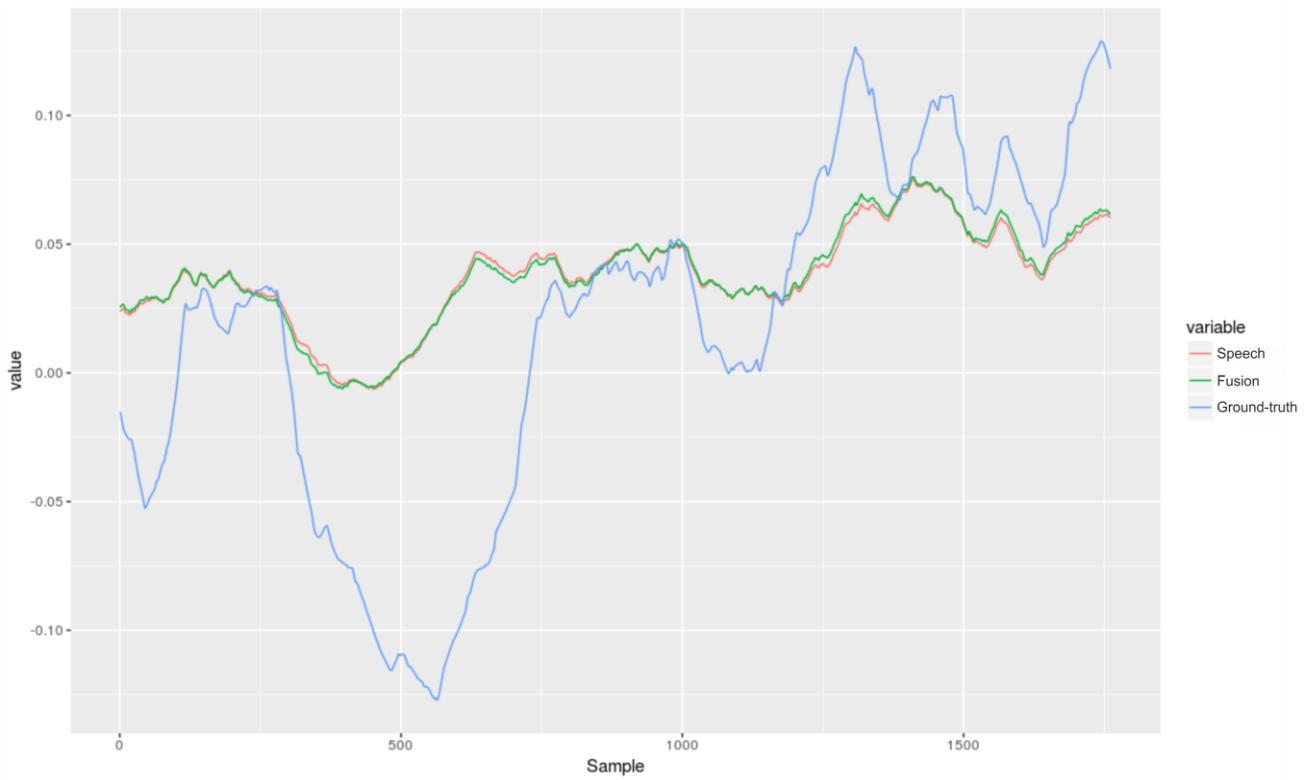

Fig. 6: Arousal prediction performance of a unimodal speech-based system compared with a bimodal speech and eye gaze feature fusion system and annotator rated ground-truth values.

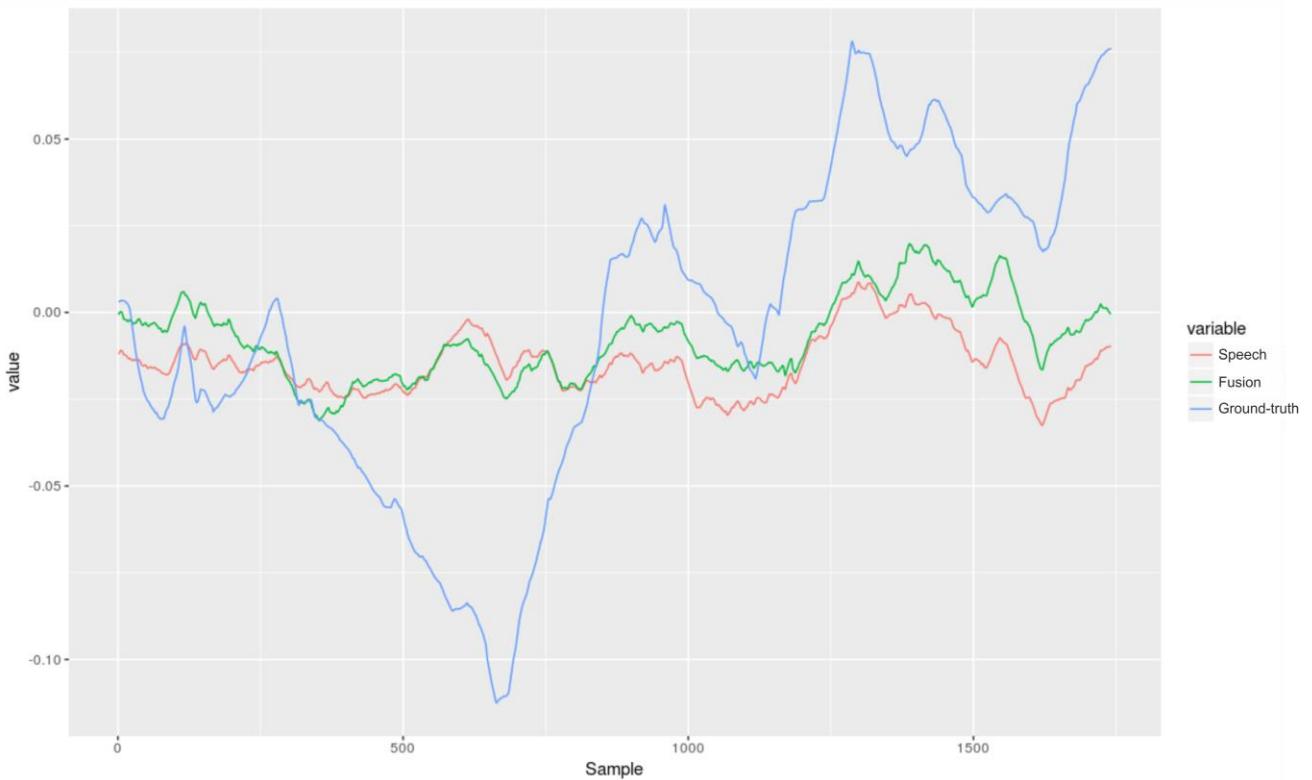

Fig. 7: Valence prediction performance of a unimodal speech-based system compared with a bimodal speech and eye gaze model fusion system and annotator rated ground-truth values.